\documentclass[ sigconf, authorversion, nonacm, balance=false]{acmart}
 \pdfoutput=1
 
\usepackage{booktabs}

\newcommand{\bits}{\{0,1\}}

\DeclareMathOperator{\GF}{GF}

\DeclareMathOperator{\trunc}{trunc}

\newcommand{\DB}{\mathcal{DB}}
\newcommand{\K}{\mathcal{K}}
\newcommand{\FP}{\mathcal{Y}}

\newcommand{\db}{\mathrm{db}}
\newcommand{\fp}{y}

\DeclareMathOperator{\init}{init}
\DeclareMathOperator{\test}{test}

\DeclareMathOperator{\Coll}{Coll}
\DeclareMathOperator{\Disc}{Disc}

\newcommand{\univ}{\mathcal{U}}

\usepackage[operators,sets]{cryptocode}
\newcommand{\randsel}{\sample}

\newtheorem{game}{Game}

\usepackage{wasysym}

\newcommand{\none}{{--}}
\newcommand{\half}{\Leftcircle}
\newcommand{\full}{$\Circle$}

\usepackage{tikz}
\usetikzlibrary{calc, matrix, fit, shapes, shapes.misc, patterns, decorations.markings, decorations.pathreplacing, positioning}
\tikzset{>=latex, shorten >=3pt, shorten <=3pt}
\tikzstyle{every picture}+=[remember picture]

\begin{document}

\title{Protecting Onion Service Users Against Phishing}

\author{Benjamin Güldenring}
\orcid{0009-0000-3333-2739}
\affiliation{
  \institution{Freie Universität Berlin}
  \city{Berlin}
  \country{Germany}}
\email{benjamin.gueldenring@fu-berlin.de}
\authornote{Corresponding author.}

\author{Volker Roth}
\affiliation{
  \institution{Freie Universität Berlin}
  \city{Berlin}
  \country{Germany}}
\email{volker.roth@fu-berlin.de}

\begin{abstract}
\bfseries
Phishing websites are a common phenomenon among Tor onion services, and 
phishers exploit that it is tremendously difficult to distinguish phishing from authentic onion domain names.
Operators of onion services devised several strategies to protect their users against phishing. 
But as we show in this work, none protect users against phishing without producing traces about visited services - something that particularly vulnerable users might want to avoid. 
In search of a solution we review prior research addressing this problem, and find that only two known approaches, hash visualization and PAKE, are capable of solving this problem.

Hash visualization requires users to recognize large hash values.  In order to make hash visualization more practical we design a novel mechanism called recognizer, which substantially reduces the amount of information that users must recognize. 
We analyze the security and privacy properties of our system formally, and report on our prototype implementation as a browser extension for the Tor web browser.

\end{abstract}

\maketitle



\section{Introduction}
The Tor~\cite{dingledine_2004_Tor} network provides internet users with a low-latency and anonymous internet experience.  Onion services are network services within the Tor network with unique security properties, such as receiver anonymity and censorship resistance.  These properties make them attractive for  journalists, human rights defenders and political activists, but also to users and merchants on market places for illicit goods.  The latter are among the more well-known~\cite{yoon_2019_Doppelgangers}
targets of phishing in the Tor network, but the Guardians' whistleblowing contact service has previously been targeted as well~\cite{lawrence_2019_Phishing}.

Phishing on onion services has been increasingly studied~\cite{barr-smith_2020_Phishing,steinebach_2021_Phishing,winter_2018_How,yoon_2019_Doppelgangers}.  Phishers exploit a central weak point in the design of onion services: onion service domains are a textual representation of a services' public key, encoded as 56 random-looking characters.  
This  makes it impossible to remember the domain names of visited services.  Human readable domain names currently exist for registered services running the SecureDrop software~\cite{onionnames}, but registration is not open for any and all other services. 

So Tor users rely on other strategies to remember the domain names of onion services, like bookmarking of domains, or relying on online directories such as Wikis~\cite{winter_2018_How}.  Bookmarking domains produces traces of visited onion services on users' computers -- something the Tor browser otherwise  attempts to prevent~\cite{TorDesign_2018}, and something that particularly vulnerable users avoid: 
In a time when onion addresses were much shorter (16 characters), a survey of Tor users found that 9\% of respondents
memorized an onion address to avoid bookmarking~\cite{winter_2018_How}.  

Online directories leave users vulnerable to Wiki poisoning attacks, where attackers replace authentic domain names with their phishing domains.   
The need to protect oneself against phishing, and the need of not leaving traces appear to contradict one another, leaving these users between a rock and a hard place.
Onion service operators deploy solutions in order to address phishing, but as we analyze in this work (\S~\ref{sec:countermeasures}), none of them solves the onion service users' conundrum.

This work takes several steps towards a solution.  We first
\begin{enumerate}
	\item[(1)] outline requirements that phishing protection mechanisms for onion service users must satisfy  (\S~\ref{sec:requirements}).  
\end{enumerate}
We base our requirements on design goals of the Tor browser and a realistic threat model. We then 
expand previous work
 by Yoon et. al.~\cite{yoon_2019_Doppelgangers} and Wang et.\ al.~\cite{wang_2024_Analysis} and

\begin{enumerate}
	\item[(2)] systematically explore phishing countermeasures that onion services currently employ, 
	and show that neither of them satisfy our security and privacy requirements (\S~\ref{sec:countermeasures}).
\end{enumerate}
We then review previous academic work (\S~\ref{sec:academicCountermeasures}) and find that only PAKE and hash visualization achieve our requirements, whereas only hash visualization does not require substantial changes to both web browsers and onion sites. 

Hash visualization requires users to recognize one large hash value (e.g. 256 bit) for each site they would like to visit.  We improve upon that by
\begin{enumerate}
	\item[(3)] designing a novel mechanism, called recognizer, that maps multiple large hash values to one much shorter value, e.g. 21 bit (\S~\ref{sec:mechanism}).
\end{enumerate}

A recognizer is a set-membership data structure.  
Recognizers are thus related to approximate membership query data structures (AMQ),  like bloom filters, whose security have received increased attention over the past few years~\cite{naor_2015_Bloom,naor_2019_Bloom,clayton_2019_Probabilistic,filic_2022_Adversarial}.  But recognizers differ in two major properties: they rely on a short password to hide the set of stored elements in an information theoretical manner, and they are allowed to output a short bit string rather than a single bit indicating set-membership.

The length of both password and fingerprint in our solution is small: for example, a configuration that 
recognizes two onion domains requires to recall a 21 bit password 
and to recognize a 21 bit fingerprint, while offering a security level of $10^{-4}$ against 
a series of 100  phishing attacks.

We conclude by describing a prototype implementation of our approach as a browser extension for the Tor web browser (\S~\ref{sec:impl}).  A browser extension that helps to detect phishing
pages was previously suggested Winter et.\ al.~\cite{winter_2018_How}, but to the best of our knowledge, ours is the first that does not produce traces of visited sites on users' devices.

\section{Threat Model and Requirements}
\label{sec:requirements}
We now motivate and outline our threat model.  
We consider a Tor user Alice who wishes to visit a set of onion services. 
Alice is concerned about two attack scenarios:  Phishing, and disclosure.

In the phishing scenario,  an adversary attempts to impersonate an onion service that 
Alice visited earlier.  There are multiple ways how an adversary might attempt phishing attacks.
Using the terminology of Barr-Smith~\cite{barr-smith_2020_Phishing}, we discern between unsophisticated
and sophisticated phishing attacks.  In sophisticated attacks, attackers set up a reverse proxy that
allows them to mimic the functionality of their target site, effectively performing a man-in-the-middle attack.  
Parts of the site can be changed and manipulated, for example, by replacing cryptocurrency addresses.  
Unsophisticated phishing attacks instead attempt to impersonate their targets without interacting with 
the original site.  Our security goal against phishing is that Alice is able to recognize when a phishing attack
is happening with a sufficiently high probability.  
Note that we use the term phishing in a more general sense than usual:  it covers every case of impersonation rather than the sole capturing of credentials.

In the disclosure scenario, an adversary captures Alices' device and
attempts to find out which onion services Alice is interested in.  We assume that a 
disclosure attack happens only once, and that Alice recognizes it when it happens.  
For example, because her device is stolen, seized by authorities or briefly taken 
away at the airport.  
Our security goal against disclosure is that the adversary should 
not be able to decide which onion services Alice previously visited.  

For the disclosure scenario we assume that adversaries are capable of breaking Alice's passwords by brute force, 
by invasive attacks on secrets 
stored in hardware or by undocumented means of access to secrets that protect 
data at rest.

While ruling out secure hardware may seem excessive, we feel that recent 
high-profile attacks on secure hardware 
~\cite{andzakovic_2019_Extracting,windknown_2020_sep,borrello_2022_AEPIC,nakashima_2021_FBI}
sufficiently motivate this decision. 
It might also seem excessive to consider adversaries that are capable of breaking passwords.  However, remembering sufficiently strong passwords that withstand brute force attacks 
is a difficult task, and we seek a solution that works without this assumption. 
Our mechanism in~\S\ref{sec:mechanism} achieves this by providing information
theoretic security. 

In summary, our primary requirements are protection against:
\begin{description}
	\item[Phishing] Alice is able to recognize when a phishing attack is happening with a sufficiently high probability.  
	\item[Disclosure] An adversary inspecting Alice's device can not decide which onion services Alice previously visited.
\end{description}
We have two additional requirements.  
Though neither the original paper~\cite{dingledine_2004_Tor} nor current design documents highlight that onion services protect against censorship, the Tor project writes\footnote{\url{https://tb-manual.torproject.org/onion-services/}, Accessed: 2023-09-12.} that
\begin{quote}
Onion services’ location and IP address are hidden, making it difficult for adversaries to censor them or identify their operators.
\end{quote}
We thus argue that a protection mechanism should not be susceptible to censorship attempts, and formulate resistance to
\begin{description}
	\item[Censorship] A protection mechanism against phishing should not be susceptible to censorship by governments, private companies or other organizations.
\end{description}
as another requirement. This, in particular, rules out mechanisms that rely on a centralized trusted third party.  
Last, onion services might get hacked, taken over, or their data exposed.  We require that this does not affect a users' security against disclosure.  
\begin{description}
	\item[Independence] An onion service operator should not be able to jeopardize a users' security against disclosure.
\end{description}
We remark that this also protects the onion service operator:  it reduces the incentive to attack the service.  

As we will elaborate in the upcoming sections it is the unique combination of these requirements that makes this problem challenging.


\section{Established countermeasures}
\label{sec:countermeasures}
Before designing our own countermeasure we wanted to find out how onion services currently attempt to protect
their users against phishing, and if they have any detectable countermeasures in place.  This leads us to
our first research question.
\begin{itemize}
	\item[RQ 1] Which countermeasures do onion services use to protect their users against phishing?
\end{itemize}
We answer this research question in \S\ref{sec:onionresults}.  Having identified several
countermeasures we want to evaluate how well they protect users against phishing, which leads us to
our second research question, which we answer in \S\ref{sec:onioneval}.
\begin{itemize}
	\item[RQ 2] How effective are the phishing countermeasures in use by onion services?
\end{itemize}
We now continue to describe our method and data collection.

\subsection{Data Collection}
We manually collected and analyzed phishing countermeasures that we found on onion services.  
To do so we extracted all onion links from the link directories dark.fail and darknetlive.com 
in September 2023, and subsequently analyzed onion services from September through November
the same year.

According to dark.fail's self presentation it curates a list of presumably authentic 
domains for onion services.  The site also claims responsibility for the Onion Mirror
Guidelines (described in \S\ref{sec:omg}) to thwart phishing, which we have found 
to have been adopted by a fraction of onion services.  This made it a suitable candidate
for our study.
In order to extend our sample of onion domains we extracted all onion links from darknetlive.com/onions/.  
Note that we only aim for the discovery of established phishing mitigation strategies and not for a representative sample of onion domains.

In total we extracted 217 unique onion domains.  We removed 37 that
could not be resolved for having a v2 onion domain, 4 whose services were apparently 
seized by authorities, 72 that would not respond on either port 80 or port 443, 3 
that were mirrors of another domain in our list, and 1 that only presented the notice that it closed down.  The resulting list counted 100 unique sites.

We connected to each domain on port 80 and port 443 and, when available, 
downloaded the files on paths \texttt{/canary.txt}, \texttt{/pgp.txt}, \texttt{/mirrors.txt} and
\texttt{/related.txt}.  We explain the purpose of these files in~\S\ref{sec:omg}.  Some 
services returned 
Captchas\footnote{CAPTCHA~\cite{ahn_2003_CAPTCHA} stands for ``Completely Automated Public Turing test to tell Computers and Humans Apart''.  For cosmetic reasons we write Captcha.}
intended to prevent denial of service attacks instead of these files.
For these we visited the paths manually in order to download them. 

We then visited each of the 100 sites manually and noted any visible anti-phishing 
mechanisms we could identify.  
We created accounts at every site where we could register (42), and
searched the settings pages for further phishing protection mechanisms.
When the site supported such mechanisms, such as two-factor authentication, we configured
and enabled them.  At 10 additional services that supported registration we failed to register
accounts.

Some onion services linked to Clearnet\footnote{We call websites reachable from outside the Tor network Clearnet websites.}
 web sites that distribute authentic onion 
domain names for this service.  For each onion service, we manually searched the service
for references to such Clearnet web sites, which we also visited.

\subsection{Results}
\label{sec:onionresults}
The major categories in our dataset were services related to cryptocurrencies (12) and technology (11), anonymous marketplaces (11), forums (8), or email providers and link directories (each 7).  We provide a full list of categories in Table~\ref{table:categories2} in \S\ref{sec:appendixdata}.  

We now explain our findings.  We identified the following mechanisms that we classify
as phishing countermeasures:
\begin{enumerate}
	\item TLS certificates
	\item Clearnet Trust Anchors\footnote{
	\label{fn:yoon}These items were previously also reported briefly by Yoon 
	et.\ al.~\cite{yoon_2019_Doppelgangers}.
	}
	\item Onion Mirror Guidelines
	\item Anti-Phishing Captchas~\footnote{
	\label{fn:wang}These items were previously also reported briefly by 
	Wang et.\ al.~\cite{wang_2024_Analysis}.}
	\item Two-Factor authentication~\textsuperscript{\ref{fn:yoon}}
	\item Address Verification Service~\textsuperscript{\ref{fn:yoon}}
	\item Login Customization~\textsuperscript{\ref{fn:wang}}
\end{enumerate}
The intent behind some of these mechanisms may not be protection against phishing.  We still count them as countermeasures against phishing when they serve the purpose.  For example: publishing the onion domain on a Clearnet web page provides discoverability, but also provides a way to
verify if a given onion domain is authentic.  We now explain these mechanisms in detail.

\subsubsection{TLS Certificates}  17 websites in our sample provided TLS certificates.  TLS certificates can vouch for the identity of an onion service in two ways.  Either as extended validation (EV) certificate that indicates the company name operating the site,
which can be acquired since 2015~\cite{CAB144}.
Or as domain validated (DV) certificate, which the certificate authority Harica issues since 2021.\footnote{\url{https://blog.torproject.org/tls-certificate-for-onion-site/}. Accessed: 2023-09-12.}.  In our dataset, 7 services used a EV and 7 a DV certificate.  Three additional services used a DV certificate that did not contain an onion domain but a Clearnet domain,
resulting in a certificate warning.

\subsubsection{Clearnet Trust Anchors} 
The vast majority (86\%) of onion services in our data used some form of Clearnet website to distribute
links to the onion domain.  We call these Clearnet Trust Anchors, short Anchors.  
We found links to be distributed on Anchors in three ways:  the page sets a dedicated 
HTTP header or meta tag~\cite{george_2018_Onion} (54), contains a link (68), or redirects (1).  15\% of the onion services with Anchors
provided only limited functionality, if at all, on their Clearnet page, while 82\% provide the same
functionality.  The remaining onion services rely on Anchors on 3rd party websites, 
to which they refer on their onion service.
Links distributed on Anchors are sometimes also signed with a PGP key.  

\subsubsection{Onion Mirror Guidelines} 
\label{sec:omg}
The Onion Mirror Guidelines~\cite{darkFailOmg} (OMG)
are a set of guidelines advertised by the dark.fail website to thwart phishing.  
Yoon et.\ al.~\cite{yoon_2019_Doppelgangers} previously observed that many onion services have multiple domain names.  The OMG describe, among other things, that onion services shall put a PGP signed document containing official mirror domain names at the path \texttt{/mirrors.txt}, and the accompanying PGP public key at \texttt{/pgp.txt} on their website.  Users are expected to store the PGP public key and use it to verify domain names published in the future.
The additional files \texttt{/related.txt} and \texttt{/canary.txt} 
provide a list of related onion services, and a warrant canary.\footnote{See \url{https://www.eff.org/deeplinks/2014/04/warrant-canary-faq}. Accessed 2023-11-20.}
In our dataset 20 onion services provided both the \texttt{mirrors.txt} and \texttt{pgp.txt} file.

\subsubsection{Anti-Phishing Captchas}  
\label{sec:apcaptchas}
Six marketplaces in our dataset used one particular type of Captcha to defend against phishing attacks.
Traces of the idea to use Captchas in this way can be found in the literature~\cite{ahmad_2018_Detecting,ferraropetrillo_2014_Design,saklikar_2008_Public}, and we discuss these in~\S\ref{sec:obc}.  We call them Anti-Phishing Captchas.

The idea of these Captchas is to combine the Captcha-solving task with the task of verifying the domain name.  The assumption is that man-in-the middle attackers can neither solve nor manipulate the Captcha in a meaningful way, and thus the Captcha serves as a trusted path (see also~\ref{sec:trustedpath}) between the onion service and the user.  This trusted path is then
used to communicate the authentic domain name to the user, which they are expected to verify.

We discern two types of Anti-Phishing Captchas.  In the first, the Captcha contains parts of the domain name, and users are expected to verify the domain name in the address bar.  Figure~\ref{fig:phishsimple} shows one example of this type of Captcha.
Solving a Captcha of the first type does not require actual verification of the domain name.  The second type of Captcha does that.  In these, users are required to read the domain name as presented in the URL bar in order to solve the Captcha.  Figures~\ref{fig:phishpart} gives one example where users have to enter parts of the URL taken from the URL bar.  

\subsubsection{Login Customization}
\label{sec:onionLogin}
One onion services allowed users to register an \emph{anti-phishing-phrase} that the service is supposed to show
after login.  This approach is very similar to Sitekey, which we discuss in~\S\ref{sec:sitekey}.
A second service allowed to register a \emph{secret phrase}, but did not provide a description on its meaning.  In our tests both services failed to show the registered phrase upon login.

\subsubsection{Two-Factor Authentication}
Some onion services employed two-factor authentication (2FA), which we count as anti-phishing measure.  We found 6 websites that used TOTP, and 10 that used PGP-based 2FA.  We  set up PGP 2FA on 8 services, the setup at two services failed for reasons unknown to us.

Setup always required to upload a PGP public key to the service.  Subsequent logins required to decrypt a challenge encrypted with this key.  The decryption then uncovered a login token that is required in order to proceed.
The details were implemented differently on each website, and we encountered the following strategies:
The decrypted challenge
\begin{itemize}
	\item  is unsigned and contains the login token (4 times).
	\item  is unsigned and contains the login token and authentic domain name (1 time).
	\item  contains a PGP-signed authentic domain name, and an unsigned login-URL containing the login token (1 time).
	\item  is PGP-signed, containing the login token and authentic domain name (2 times).
\end{itemize}
In addition to TOTP one service supported two-factor authentication based on SMS and security keys (Facebook).

\subsubsection{Address Verification Service}
One onion service provided an address verification service, where users can test if a given onion 
domain is a trusted mirror of the service.

\begin{figure}
\centering
\includegraphics[width=0.3\textwidth]{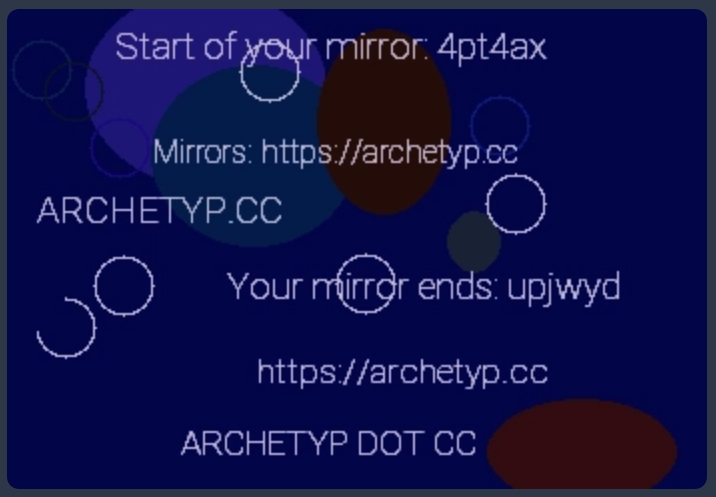}
\caption{Shows a phishing protection mechanism that combines solving a Captcha with the verification of a domain name.  
The Captcha serves as trusted path between the server and the user, communicating the correct domain name.
\protect\footnotemark
}
\label{fig:phishsimple}
\end{figure}

\footnotetext{
\label{fn:wangimages}Compare parallel work by Wang et.\ al.~\cite{wang_2024_Analysis}.
}

\begin{figure}
\centering
\includegraphics[width=0.3\textwidth]{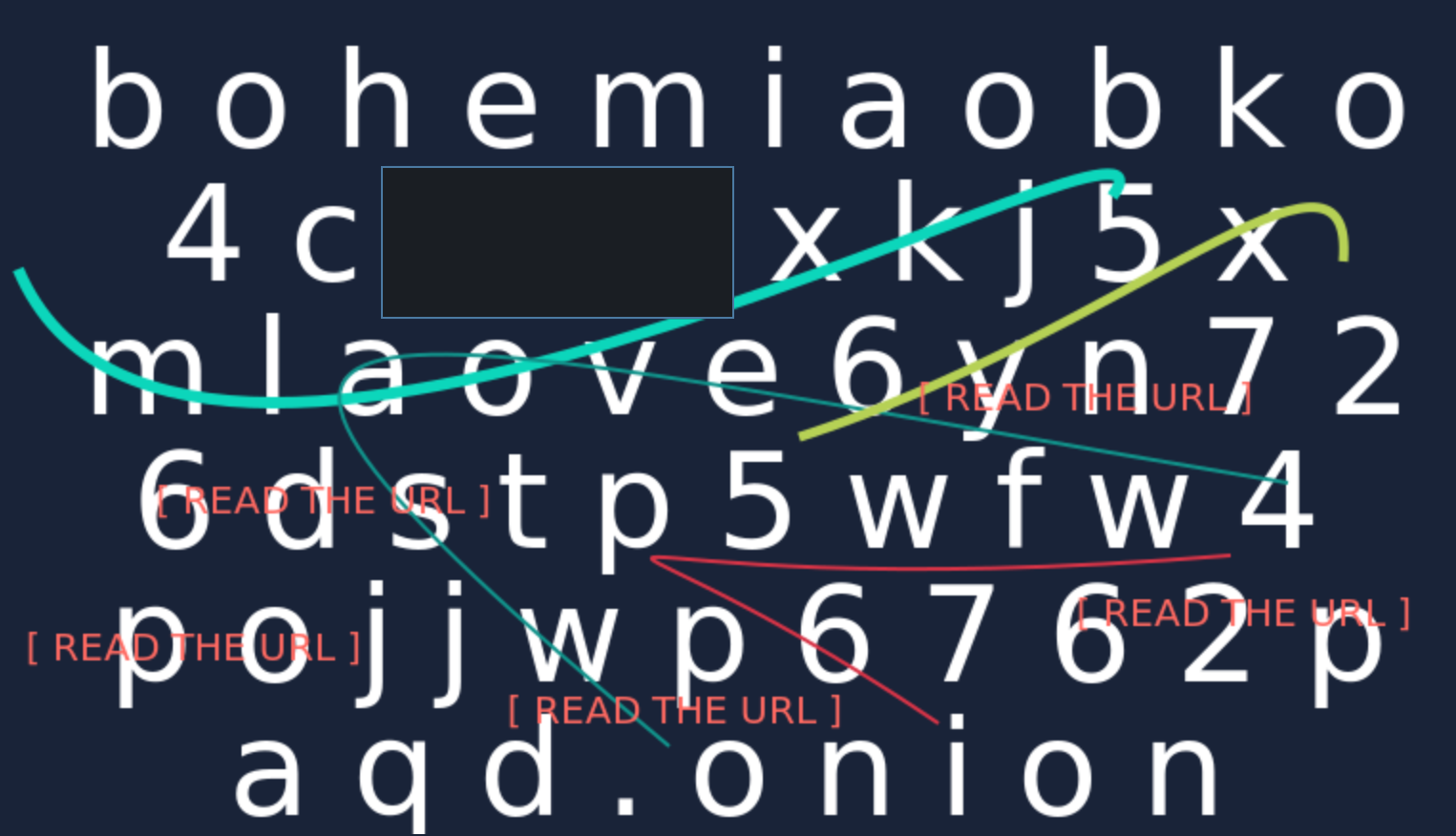}
\caption{Another phishing protection mechanism based on a Captcha.  Users enter the missing characters of the domain name  as shown in the URL bar, thereby verifying parts of it.\textsuperscript{
	\ref{fn:wangimages}
}
}
\label{fig:phishpart}
\end{figure}

\subsection{Security Evaluation}
\label{sec:onioneval}
We now evaluate these countermeasures with respect to our requirements: protection against phishing, disclosure attacks, censorship, and independence.  
Table~\ref{tab:meval} summarizes the following discussion.

We apply the following grading system:  A full circle \full{} indicates a solution fully satisfies a property, a half circle \half{} that it only satisfies the requirement partially or that it needs further assumptions, and a dash~\none{} that it does not satisfy the requirement.

\begin{table*}
\caption{Shows phishing protection methods of onion services in our dataset, and how well they perform in our evaluation.  Legend:
 \none: requirement is not met;  \half: requirement is met under assumptions; \full: requirement is met without assumptions.}
\label{tab:meval}
\begin{tabular}{l|ccc|c}
\toprule
& \multicolumn{3}{c|}{Protection against}\\
Countermeasure & Phishing & Disclosure & Censorship & Independence \\ 
\midrule
TLS certificates &				\half 		& \full 	& \none 	& \full \\
Clearnet Trust Anchors &			&&&\\
\quad Dedicated Pages &		\half		& \full	& \half 	& \full \\
\quad Spreading Links &			\half		& \full	& \full 	& \full \\
Onion Mirror Guidelines  &		\full		& \none	& \full 	& \none \\
Login Customization &			\half		& \full	& \full 	& \full \\
Anti-Phishing Captchas &			\half		& \full	& \full 	& \full \\
Address Verification Service &		\none	& \full	& \full 	& \full \\
PGP Two-Factor Authentication	& 	\full		& \full	& \full 	& \none \\
\bottomrule
\end{tabular}
\end{table*}

\subsubsection{TLS Certificates}  
TLS certificates rely on an external root of trust: the certificate authority (CA) ecosystem. The assumption is that no CA will issue a certificate to the wrong person, which effectively turns the CA ecosystem into a net of trusted third parties.  CAs as legal entities can be pressured by governments, leaving this solution vulnerable to censorship  (Censorship: \none).
Furthermore, the CA ecosystem has systemic issues arising from a misaligned incentive structure~\cite{malchow_2018_New}.  Best practice, at the moment, is to systematically monitor certificates served by web servers for signs of misbehaving CAs~\cite{laurie_2014_Certificate}, which was recently proposed for onion services as well~\cite{dahlberg_2022_Sauteed}.  While monitoring helps to detect phishing attacks it does not prevent them (Phishing: \half).  TLS certificates don't store anything on clients, and thus provide security against disclosure (Disclosure: \full; Independence: \full).

\subsubsection{Clearnet Trust Anchors}  
Clearnet Trust Anchors have similar limitations as TLS certificates.  Domain names can be spread on dedicated Clearnet web pages, or in forums and link directories.

Dedicated web pages allow users to determine which link is authentic and which one is not.  But the identity of the dedicated page relies on third parties: the DNS infrastructure, or the CA ecosystem verifying this domain.  This makes them susceptible to censorship and phishing (Phishing: \half; Censorship: \half), while entailing no risk against disclosure (Disclosure: \full; Independence:~\full).

Domain names spread in forums or directories are either susceptible to wiki poisoning, or require moderators that filter out phishing links, effectively acting as trusted third parties.  It is difficult to prevent domain names from being spread on platforms, though (Phishing: \half;  Censorship: \full; Disclosure: \full; Independence:~\full).

\subsubsection{Onion Mirror Guidelines}  
Onion services following the Onion Mirror Guidelines provide a list of PGP signed mirrors at a standard location on their web site.  The public key is distributed on the same web site at a different location.  Downloading the key every time one wants to verify the list of mirrors renders the signature useless.  Users are thus required to store the key that was generated by service operators (Disclosure: \none; Independence: \none).  On the upside this approach helps to securely recognize the same site at the same and different onion addresses (Phishing: \full), and
does not require a trusted third party (Censorship: \full).

\subsubsection{Login Customization}
\label{sec:onionLoginCustom}
Login customization only helps against unsophisticated (see \S\ref{sec:requirements}) phishing attacks.  Sophisticated phishing pages that run man-in-the-middle attacks will relay the login message.  Research on security indicators has also shown that users do not notice the absence of these messages~\cite{schechter_2007_Emperor}  
(Phishing: \half).  Login customization does neither require users to store anything, nor is it susceptible to censorship (Disclosure: \full; Independence: \full; Censorship: \full).

\subsubsection{Anti-Phishing Captchas}  
\label{sec:onionCaptcha}
Anti-Phishing Captchas are neither susceptible to censorship, nor do they require users to store data.
Their security against phishing hinges on three assumptions:  
\begin{enumerate}
	\item that an adversary can not solve the Captcha, 
	\item that the domain name embedded in the Captcha can not be manipulated, and
	\item that users actually verify the domain name.
\end{enumerate}
We will concentrate on the latter assumptions first.  Captchas that do not incorporate the domain name 
in solving the Captcha, like those in Figure~\ref{fig:phishsimple}, fail to ensure that users actually verify the domain name.  We know that users tend to skip unnecessary steps even when doing so is for their own safety.  In other words, solving these Captchas is not a conditioned-safe ceremony~\cite{karlof_2009_Conditionedsafe}.

Captchas that enforce verifying the domain name, like the one in Figure~\ref{fig:phishpart}, are however much more difficult to design to resist image recognition and manipulation.  
The problem is that the adversary \emph{knows} the correct domain name, and only has to decide how the challenge is formed.
Captcha systems from the literature, discussed in~\S\ref{sec:obc}, avoid this problem using randomization, at the expense of requiring modifications to the web browser. 
We leave practical attacks against the Captchas we found for future work, but we argue that it seems difficult to design them in such a way that they resist attacks using modern image recognition capabilities. 

Captchas can also not protect against human-in-the-middle attackers and static phishing pages (Phishing: \half).  They are however not susceptible to censorship nor disclosure (Censorship: \full; Disclosure: \full; Independence: \full).

\subsubsection{Address Verification Service}
Address verification services only protect against phishing when users know the correct onion domain where to find this service.  This might help onion service operators in discovering potential phishing pages, but offers no protection to users (,Phishing: \none; Disclosure: \full; Censorship: \full;  Independence:~\full).

\subsubsection{PGP Two-Factor authentication}
In order to use PGP-based two-factor authentication users need to register a PGP public key. 
When attackers gain access to a list of registered keys,
and subsequently to the private key in a disclosure attack, they can use this private key in order to 
prove that it matches a given public key.  This binds a user's security in the disclosure scenario on how well the onion service protects the public key (Independence: \none).

Man-in-the-middle attackers can relay encrypted challenges to users.  Challenges thus need to contain the authentic domain name to thwart phishing.  Furthermore, the encryption scheme needs to protect against modification of ciphertexts.  While previous versions of PGP were susceptible to ciphertext modifications~\cite{poddebniak_2018_Efail}, more recent drafts (e.g.\ \cite[\S 13.7]{wouters_2024_OpenPGP}) avoid to decrypt malleable or manipulated 
ciphertexts.
In summary, PGP two-factor authentication challenges can prevent phishing, but require users to store a PGP private key that fits a public key on the server (Phishing: \full; Disclosure: \full; Independence: \none).  PGP two-factor
authentication is not susceptible to censorship, though (Censorship: \full).

\subsection{Discussion}
We have seen that onion services utilize various phishing countermeasures.  However, none of them fully satisfies all of our
requirements.  
Solutions not requiring to store data are limited: Login messages won't protect against sophisticated phishing, Clearnet Trust
Anchors only shift the problem to finding an authentic Anchor, and Anti-Phishing Captchas can easily be defeated by humans.

Solutions that store data are vulnerable against disclosure. The discussion about OMG and 2FA highlights the problem:  users store a public key of services (OMG), or services store a public key of users (PGP 2FA).  On either side users or services keep information that links them.  As we will see in \S\ref{sec:dyn2fa} this problem also extends to other forms of 2FA.



\begin{table*}
\caption{Summarizes our discussion in \S\ref{sec:academicCountermeasures}
 software-based interventions 
designed to prevent phishing attacks, and how well they satisfy our requirements from~\S\ref{sec:requirements}.
Legend:
 \none: this property is not fulfilled; \half: this property is fulfilled under assumptions; \full: this property is fulfilled without assumptions; ?: open question.
}
\label{table:relatedwork}
\begin{tabular}{l|ccc|cc}
\toprule

& \multicolumn{3}{c|}{Protection against}&\\
Countermeasure & Phishing & Disclosure & Censorship & Independent & References \\ 
\midrule
Identity Indicators           		& \none 	& \full 	& \full 	& \full 	& 
  	e.g. \cite{lin_2011_Does,miyamoto_2014_EyeBit} \\
Sitekey \& Personalization         & \none 	& \full 	& \full 	& \full 	& 
	\cite{bankofamerica_2005_Bank,iacono_2014_UIDressing,herzberg_2013_Forcing} \\ 
  Static 2FA                			& \none 	& \full 	& \full 	& \none 	& 
	\cite{rfc6238} \\
 Assisted Decision Making       & \none & \full & \full & \full &
  \cite{ronda_2008_Itrustpage,likarish_2009_BayeShield} \\
Sauteed Onions            & \none & \none & \full & \full & 
  \cite{dahlberg_2022_Sauteed} \\

Onion DNS               & \none & \full & \none & \full & 
  \cite{scaife_2018_OnionDNS} \\
Tor Proposal 194, Mnemonic              & \none & \full & \full & \full & 
  \cite{sai_2012_Mnemonic} \\

\hline

Trusted Path Captchas           & \half & \full & \full & \full   &
\cite{ahmad_2018_Detecting,ferraropetrillo_2014_Design,saklikar_2008_Public,urban_2017_Riddle} \\ 
 Allow- and Deny-Lists       & \half & \full & \none & \full &
  e.g.\ \cite{google_2023_Google} \\ 
 Stateless Heuristics        & \half & \full & \full & \full &
  e.g.\ \cite{chou_2004_Clientside} \\  
 Stateful Heuristics       & \half & \none & \full & \full &
  \cite{chou_2004_Clientside,ardi_2016_AuntieTuna} \\ 

Onion Names               & \half & \full & \none & \full & 
  \cite{onionnames} \\

Password Generators           & \half & \full & \full & \full  &
  e.g. \cite{gabber_1997_How,ross_2005_Stronger,halderman_2005_Convenient,yee_2006_Passpet} \\
Onion Name System           & \half & \full & \full & \full & 
  \cite{victors_2017_Onion} \\

\hline

Client Secrets Stored in Cookies	& \full 	& \none 	& \full 	& \full 	& 
	\cite{juels_2006_Cache,hart_2011_PhorceField,braun_2014_PhishSafe} \\
BeamAuth					& \full 	& \none 	& \full 	& \full 	& 
	\cite{adida_2007_Beamauth} \\
Credentials Tracking        		& \full & \none & \full & \full &
  \cite{kirda_2006_Protecting,yue_2012_Preventing,wu_2006_Web} \\

Password Managers           & \full & \none & \full & \full &
\\
Theft-Resistant Password Managers     & \full & \none & \full & \full &
  \cite{bojinov_2010_Kamouflage,wolf_2012_IMobileSitter, chatterjee_2015_CrackingResistant,guldenring_2015_Knock} \\
Petnames              & \full & \none & \full & \full &
  e.g. \cite{close_2006_Petname,yee_2006_Passpet,rfc3709,josang_2001_Authentication,close_2004_Trust,herzberg_2008_Security,rivest_1996_SDSIa} \\

Let's authenticate			& \full 	& \half 	& \full 	& \half 	& 
	\cite{conners_2022_Let} \\
	
WebAuthn             			& \full 	& \half 	& \full 	& \half 	& 
	\cite{Webauthn} \\

Slowing Down Password Cracking        & \full & \half & \full & \full &  e.g. \cite{canetti_2006_Mitigating,daher_2008_POSH,rfc2898,blocki_2013_GOTCHA,manber_1996_Simple,bai_2022_Costasymmetric}\\

\hline

  BBMA					& \full 	& \full 	& \full 	& \none 	& 
	\cite{gajek_2008_Provably} \\

\hline

\textbf{PAKE protocols}            &  \full 	& \full 	& \full 	& \full 	& 
  e.g.~\cite{jakobsson_2007_Delayed,dhamija_2005_Battle,wu_1998_Secure,jarecki_2018_OPAQUE} \\

\textbf{Hash Visualization}             & ? & \full & \full & \full & 
  e.g. \cite{perrig_1999_Hash,tay_2004_Visual,dabas_2005_Browser,markham_2005_Phishing} \\
\bottomrule
\end{tabular}
\end{table*}

\section{Known Countermeasures Against Phishing}
\label{sec:mainAcademicCountermeasures}

We discuss a more comprehensive list of known countermeasures from the literature in Appendix~\S\ref{sec:academicCountermeasures} and only provide a summary of our results here.  
Table~\ref{table:relatedwork} gives an overview of this discussion.  Most mechanisms that protect against sophisticated and unsophisticated phishing attacks do so by storing information about the web site on clients (third group in Table~\ref{table:relatedwork}), or by storing a public/private key combination on clients and websites (BBMA, fourth group).  To the best of our knowledge, the only approaches that  fulfill our requirements without doing so are PAKE and hash visualization.

In PAKE (see also \S\ref{sec:pake}), the web browser and server run a cryptographic protocol computing a shared session secret that is authenticated by a previously shared secret between browser and server, such as the user's password.  Modern PAKE protocols are designed such that an active man-in-the-middle adversary can't compute the password nor the session secret, and are thus resistant to sophisticated phishing attacks.  PAKE must be implemented in browsers and servers, and is currently not implemented in modern web browsers.

PAKE requires a previously shared secret between user and website.  
Though not one of our requirements (compare \S\ref{sec:requirements}), it is beneficial to have a solution that works without this requirement.  Onion services such as news sites or search engines do not require registration, and adding such a step solely for the purpose of protection against impersonation seems excessive.  
Hash visualization provides such a solution, and we discuss this next in more detail.

\subsection{Hash visualization}
\label{sec:keyvis}

Hash visualization, calculates images from public key information, called visual hash or visual fingerprint, and users are expected to recognize when being shown the same or different images.

Originally proposed by Perrig and Song in the context of root key verification~\cite{perrig_1999_Hash}, hash visualization was later also suggested to visualize web site certificates~\cite{tay_2004_Visual,dabas_2005_Browser}.  While we are not aware of any widespread use in browsers, one variant was adopted in OpenSSH 5.1, and later dubbed drunken bishop~\cite{loss_2009_Drunken}.   Hash visualization was  also discussed\footnote{\url{https://lists.torproject.org/pipermail/tor-dev/2015-August/009302.html}, Accessed 2023-11-29.} among Tor developers in order to visualize onion domains, but eventually not adopted.

While there has been research on the comparison of keys using different visualizations 
(\cite{dechand_2016_Empirical,tan_2017_Can,azimpourkivi_2020_Human,turner_2023_Effect}), 
research on the recognition of key visualizations in the context of phishing is sparse.
Wiese used a simulation-based approach in order to analyze the recognition of faces in the context of email phishing~\cite{wiese_2023_Phish}. 
But in summary, the effectiveness of hash visualization against phishing is an open question (Phishing:~?).

Hash visualization is limited by the requirement to encode a large number of bit (e.g. 256). 
Markham suggested to add a small 12 bit hash value of a domain name to the domain~\cite{markham_2005_Phishing} in order to thwart typo-squatting.  But
without further safeguards, the 56 characters in an onion domain provide ample 
opportunity to brute-force fitting pre-images of a 12 bit hash value.

The mechanism introduced in the following section addresses this limitation:  it produces much smaller outputs while still providing a suitable notion of collision resistance.
The calculation also allows for deliberate collisions:  it produces the same output for a previously selected set of inputs.  This allows users to recognize only one visual hash for multiple onion sites instead of multiple distinct ones.


\newpage
\section{Secure Recognizers}
\label{sec:mechanism}

\begin{figure}
\centering
\includegraphics[width=0.25\textwidth]{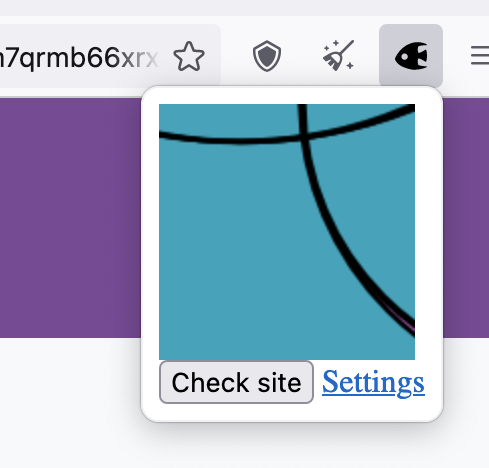}
\caption{Our browser extension prototype calculates a domains' recognizer fingerprint and shows it as visual hash.}
\label{fig:extensioncheck}
\end{figure}

Before we describe the mechanism behind our approach we briefly explain how it works for users.
Our approach builds on the idea of hash visualization (see \S\ref{sec:keyvis}).  Onion domain names are transformed into visual hashes in such a way that hashes are short, but it is still difficult for adversaries to produce collisions.  Furthermore, users can select multiple onion domain names to produce the same hash. 

On a high level this works as follows.  Users first select a series of onion domains, and receive a password and reference visual hash in return.   Figure~\ref{fig:extensioncheck} shows such a visual hash in our prototype browser extension, which we discuss in~\S\ref{sec:impl} in more detail.  Users check a website by entering their password and recognizing if the shown visual hash matches the reference visual hash shown earlier.

At the heart of our solution is a construction that we call recognizer:
A recognizer is a probabilistic data structure similar to an approximate membership query data structure (AMQ).  Like an AMQ, a recognizer stores a set of elements and allows to test for membership of elements in this set.  
But there are two major differences between AMQs and recognizers.  First, recognizers output a string of bits, the fingerprint, whereas AMQs only output a single bit indicating set membership.  We will require that every element stored in a set will yield the same fingerprint.
Second, recognizers make use of a human memorable password in order to conceal the membership of elements in the set.

Note that we call the output of a recognizer fingerprint while we call the image shown to users a visual hash.  The recognizer concept is oblivious to the particular visual hash implementation used (see~\S\ref{sec:impl}).

In the following sections we formally define recognizers and their security properties.  We then give a recognizer construction and determine its security in our formalization.  Afterwards, in section~\S\ref{sec:impl}, we share how we built a browser extension for the Tor browser that uses recognizers to help users identify previously visited onion domains.

\subsection{Preliminaries}
\label{sec:prelim}
We now briefly recite some well-known facts and terms.  We will also introduce the computational and security frameworks that we use in order to analyze the security of our constructions.

We use a concrete security framework.  This means we use statements of the form \emph{any probabilistic algorithm that runs for at most time $t$ has success probability of at most $\varepsilon$} to describe the security of constructions, and will say the construction is $(t,\varepsilon)$ secure.   
For the constructions in this paper we will always have $t=\infty$, but our formalization also supports weaker solutions. 
  
Let $A$ be an algorithm.  
If $A$ is probabilistic, we write $y \leftarrow A(x)$ to denote that $y$ is obtained by running $A$ on input $x$.  When $M$ is a set, we write $x \randsel M$ to sample a value $x$ from $M$ uniformly at random.  When $A$ takes values from $M$ as input we write $A(M) := \{A(x) : x \in M\}$ to denote the set of values that $A$ outputs when receiving inputs from $M$.  Unless stated otherwise, an adversary $A$ is a probabilistic algorithm.

\paragraph{Polynomials over Finite Fields}
Let $\GF(2^n)$ denote the finite field with $2^n$ elements.
The elements of $\GF(2^n)$ are essentially bit strings of length $n$. This allows to identify bit strings with elements in $\GF(2^n)$, and we use $\GF(2^n)$ and $\bits^n$ interchangeably. 

As polynomials over a finite field, polynomials over $\GF(2^n)$ have the useful property that any
series of $k < |\GF(2^n)|$
 pairs $(x_1,y_1), \dots, (x_k,y_k)$ with distinct $x_i$ uniquely determine a polynomial $p$ of degree $k-1$ so that $p(x_i)=y_i$. 
This property makes them a useful building block in the construction of hash functions.

\begin{definition}
Let $k,m < n$ be natural numbers.  Define the family of hash functions $H$ mapping strings from $\bits^n$ to $\bits^m$ with
\[
	H := \{h_{a_0, \dots, a_{k-1}} : a_0,\dots,a_{k-1} \in \GF(2^n)\}
\] 
with $h_{a_0, \dots, a_{k-1}} : \bits^n \rightarrow \bits^m$ defined as 
\[
	h_{a_0, \dots, a_{k-1}}(x) := \trunc_m\left(a_0 + a_1 x + \dots + a_{k-1} x^{k-1}\right),
\]
where multiplication and addition are calculated in $\GF(2^n)$, and $\trunc_m(y)$ truncates $y$ it to it's $m$ least significant bit. 
\end{definition}

For $a=(a_0,\dots,a_{k-1}) \in \GF(2^n)^k$ we will also write $h_a$ instead of $h_{a_0,\dots,a_{k-1}}$.
Carter and Wegman~\cite{wegman_1981_New} first noted that this hash function family is strongly $k$-universal.  
\begin{definition}[strongly $k$-universal]
Let $H$ be a family of hash functions where $h\in H$ maps $\bits^n$ to $\bits^m$.  $H$ is called strongly $k$-universal, if for every series of $k$ pairs $(x_1,y_1), \dots, (x_k,y_k)$ with distinct $x_i$ it is
\[
	\underset{h \randsel H}{\Pr}[h(x_1) = y_1 \wedge \dots \wedge h(x_k)=y_k] = (2^{-m})^k.
\]
\end{definition}
This in particular means that for $h \randsel H$ and given any distinct values $x_1,\dots, x_k$ the random variables $h(x_1), \dots, h(x_k)$ are uniformly distributed over $\bits^m$ and independent.

\subsection{Formally Defining Recognizers}

We now define recognizers. 
A recognizer consists of two algorithms: init and test.  Init creates a new database from a set of elements taken from a universe $\univ$, while test is used for membership queries.

\begin{definition}[Recognizer]
	Let $\univ$ be a universe of items and $N < |\univ|$ be a positive integer.  
	Let further $\K, \DB, \FP$ be sets and call $\K$ the set of keys, $\mathcal{DB}$ the set of 
	databases and $\mathcal{Y}$  the set of fingerprints.
	We say $\Pi=(\univ,N,\init,\test)$ is a recognizer for $N$ items, when $\init$ is a probabilistic and $\test$ a deterministic algorithm, where
	\begin{description}
		\item[Init] $\langle \db, k, \fp \rangle \leftarrow \init(M)$ takes as input a set of 
		$N$ distinct elements 
		$M = \{x_1, \dots, x_N\}$ with $x_i \in \univ$, and returns 
		a database $\db \in \DB$,
		a key $k\in\K$, 		
		and a fingerprint $\fp \in \FP$,
		\item[Test] $\fp = \test(\db, k, x)$ takes as input 
		a database $\db\in\DB$,
		a key $k\in\K$, 
		and an item $x \in \univ$
		and returns a fingerprint $\fp\in\FP$,
	\end{description}
	and for which the following correctness condition holds:  Let $M$ be as above and $\langle \db, k, \fp \rangle \leftarrow \init(M)$.  Then we require
		\[
			\forall x \in M : \test(\db, k, x) = \fp.
		\]
	For technical reasons we require that $\K, \DB$ and $\FP$
	do not contain any elements that are never the output of $\init$ regardless of $M$.  When it is clear from context we omit the definition of $\K, \DB$ and $\FP$ for a recognizer $\Pi$ and assume they are defined implicitly.
\end{definition}

Just like AMQs the correctness property of recognizers prohibits false negatives while allowing false positives.  A Bloom filter, for example, can be seen as a recognizer that returns a fingerprint of size 1, and where $init$ and $test$ ignore the key parameter.

To make the following presentation more succinct we now introduce some terms.  Let $\Pi=(\univ, N, \init,\test)$ be a recognizer and let $R := \langle \db, k, \fp \rangle \leftarrow \init(M)$ for some set $M=\{x_1, \dots, x_N\} \subset \univ$.  We call $R$ a recognizer for $M$, and say that we stored $x$ in $R$.  For any $x'$ with $\test(\db, k, x') = \fp$ we say that $R$ recognizes $x'$.  When $x' \not\in M$ we sometimes also say that $R$ wrongly recognizes $x'$.
 
 \paragraph{Security Notions}

We now define two security notions for recognizers: security against collisions, and security against disclosure.  We model security using a game based approach, i.e., with respect to the success chance of an adversary in a security game.  

We start with security against collisions.  The collision game is supposed to model a series of phishing attacks.  The adversary is first allowed to populate a recognizer, and is then tasked to produce
a wrongly recognized element within $q$-many phishing attempts.
Each phishing attempt is modeled as access to an oracle $O^{\test}$ that calculates $\test(\db, k, \cdot)$.
This game follows the structure of the standard forgery game modeling
chosen message attacks on message authentication codes~\cite{katz_2007_Introduction}.

\begin{game}[Recognizer Collision Game]
Let $\Pi = \langle \univ,N,\init,\test \rangle$ be a recognizer for $N$ items, and let $A$ be an 
adversary. Let $q$ be a number.  The collision game $\Coll^{\Pi}_{A}(q)$ runs as follows.
\begin{enumerate}
	\item $A$ outputs a set of $N$ distinct items $M := \{x_1, \dots, x_N\} \subset \univ$.
	\item Let $\langle \db, k, \fp \rangle \leftarrow init(M)$.
	\item $A$ receives black-box access to an oracle $O^{\test}$, with 
	\[
		O^{\test}(x) = 
		\begin{cases}
		1 & \fp = \test(\db, k, x)\\
		0 & \text{otherwise}
		\end{cases}
	\]
	Let $Q$ denote the set of queries $A$ makes to $O^{\test}$.
	\item $A$ outputs an item $x$.
\end{enumerate}
The result of this experiment is $1$ if $|Q| \leq q$, $x \in Q, x\not\in M$ and $\test(\db, k, x)=\fp$, and $0$ otherwise.  We write the result as $\Coll^{\Pi}_{A}(q)$. 
\end{game}
The intention of the game is that the adversary wins whenever it queried a wrongly recognized value.  We require $A$ to output such a value at the end to make our proofs easier, but we stress that doing so poses no restriction on the capabilities of $A$.  We now define security in the collision game.

\begin{definition}
Let $\Pi = \langle \univ,N,\init,\test \rangle$ be a recognizer for $N$ items.   We say $\Pi$ is $(q,t,\varepsilon)$-secure against collisions, if for every probabilistic adversary $A$ that runs in at most time $t$ it holds that
\[
	\Pr[\Coll^{\Pi}_{A}(q) = 1] \leq \varepsilon,
\]
where the probabilities are taken over the choices of $A$ and $\Pi$.
\end{definition}

We proceed with security against disclosure.  In the disclosure game, the adversary is first allowed to output two candidate sets of items $M_0$ and $M_1$. A recognizer is initialized with one randomly chosen set, and it's database given to the adversary.  The adversaries' task is then to guess which elements have been stored in the recognizer.  
Note that the adversary does not get access to the key $k$, fingerprint $\fp$ or a test oracle. 
This game follows the structure of the standard eavesdropping game modeling chosen plaintext attacks on private key encryption~\cite{katz_2007_Introduction}.

\begin{game}[Recognizer Disclosure Game]
Let $\Pi = \langle \init, \test \rangle$ be a recognizer for $N$ items and let $A$ be an
adversary. The disclosure experiment $\Disc^{\Pi}_{A}(N)$ runs as follows.
\begin{enumerate}
	\item $A$ outputs $M_0, M_1 \subset \univ$ with $|M_0| = |M_1|=N$.
	\item Choose a random bit $b \leftarrow \{0, 1\}$
	\item Let $\langle \db, k,\fp, \rangle \leftarrow \init(M_b)$
	\item Give $\db$ to $A$.
	\item $A$ outputs a bit $b'$.
\end{enumerate}
The result of this experiment is $1$ if $b=b'$, and 0 otherwise.  We write the result as $\Disc^{\Pi}_{A}(N)$. 
\end{game}

Security in the disclosure game means that $A$ cannot decide which set was chosen.
Security must also be derived from $A$ knowing neither key $k$ nor fingerprint $\fp$, since
every other information is known to $A$.  Contrast this to the collision game, where $A$ only 
receives access to an oracle $O^{\test}$ and the database might contain additional secrets.

The key $k$ will be encoded into a password that the user has to recall, and the fingerprint $\fp$
 will be something that the user has to recognize.  The bit-length of both must thus be rather small,
and we assume that adversaries are capable of exhaustive attacks over key and fingerprint.
For this reason we do not restrict the computation time of $A$ in the following definition.
We also only consider a perfect notion, i.e., $A$ can only win with chance $0.5$.

\begin{definition}[Security against Disclosure]
Let $\Pi$ be a recognizer for $N$ items.  We say that $\Pi$ is secure against disclosure, when for all  adversaries $A$ it is
\[
	\Pr[\Disc^{\Pi}_{A}(N)=1] = 0.5.
\]
\end{definition}

Our last definition summaries both security notions.

\begin{definition}
A recognizer $\Pi$ is called $(q, t,\varepsilon)$-secure if it is $(q, t,\varepsilon)$-secure against collision, and secure against disclosure.
\end{definition}

\subsection{A Recognizer using Polynomials}
\label{sec:polyrec}
\begin{figure}
\centering
\begin{tikzpicture}[
node distance=1.2cm and 1.2cm,
draw, rectangle,
minimum width=1cm
]
	\node[draw, rectangle] (X) {$x$};
	\node[right =of X, draw, rectangle] (H) {$h_{\db}(\cdot)$};
	\node[right =of H, draw, rectangle] (TEST) {$p(\cdot)$};
	\node[right =of TEST, draw, rectangle] (Y) {$\fp$};

	\node[draw, rectangle, anchor=south, dashed] at (TEST.north) {$k$};

	\node[draw, rectangle, anchor=south, dashed] at (H.north) {$\db$};

	\draw[->] (X) -- (H);
	\draw[->] (H) -- (TEST);
	\draw[->] (TEST) -- (Y);

	\coordinate (tmp1) at ($(X) ! 0.7 ! (H)$);
	\coordinate (tmp2) at ($(H) ! 0.3 ! (TEST)$);
	
	\draw[decorate,decoration={brace,amplitude=5pt}] 
		($(tmp2) + (0, -0.5cm)$) -- ($(tmp1) + (0,-0.5cm)$)
		node[midway, below, yshift=-0.3cm] {$N+q$-universal};
		
\end{tikzpicture}
\caption{Illustrates the recognizer from Definition~\ref{def:recogdb}.  A $(N+q)$-universal hash function $h$ shortens the input of $x$ to a value $\hat{x} = h_{\db}(x)$.  The result is then fed into a polynomial $p$ that returns the same value for every stored item.}
\label{fig:serial}
\end{figure}
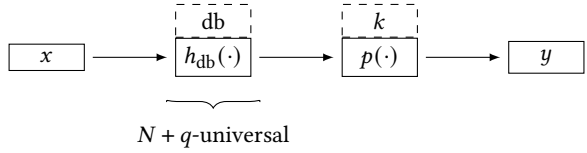

We will now describe our recognizer construction.  The following Lemma will be useful in our construction:  given $N$ numbers, it tells us how to construct a polynomial of degree $N+1$ that gives the same result on each of those numbers.

\begin{lemma}
\label{lemma:poly}
Let $x_1, \dots, x_{N} \in \GF(2^m)$ be $N<2^m$ distinct field elements.  The polynomial 
\[
	p(x) := \prod_{i=1}^{N} (x-x_i) + \prod_{i=1}^{N} x_i
\]
over $\GF(2^m)$ has the following properties:
\begin{enumerate}
	\item $p$ is of the form $p(x) = x^N + a_{N-1}x^{N-1} + \dots a_1 x$, and
	\item $p(x_i) = p(x_j)$ for all $1 \leq i,j \leq N$.
	\item\label{lemma:poly:property} $\forall x \in \GF(2^m) : p(x) = p(x_1) \Rightarrow x \in \{x_1, \dots, x_N\}$.
\end{enumerate}

\begin{proof}
We have to show three things:
\begin{enumerate}
	\item $p$ is of the form $p(x) = x^N + a_{N-1}x^{N-1} + \dots a_1 x$, and
	\item $p(x_i) = p(x_j)$ for all $1 \leq i,j \leq N$.
	\item$\forall x \in \GF(2^m) : p(x) = p(x_1) \Rightarrow x \in \{x_1, \dots, x_N\}$.
\end{enumerate}

We prove each item separately.
\begin{description}
\item[(1)] We show that $\prod_{i=1}^{N} (x-x_i)$ is of the form
\[
	x^N + b_{N-1}x^{N-1} + \dots + b_1 x + b_0,
\]
with $b_0 = \prod_{i=1}^N -x_i$.  Since $-x=x$ over $\GF(2^m)$, $\prod_{i=1}^{N} (x-x_i) + \prod_{i=1}^N x_i$ will then have the desired form.  We prove the statement by induction over $N$.  For $N=1$ the statement is true.  for $N>1$ we have
\begin{align*}
\prod_{i=1}^{N}& (x-x_i) = (x-x_N) \cdot \prod_{i=1}^{N-1} (x-x_i)\\
&= (x-x_N) (x^{N-1} + c_{N-2}x^{N-2} + \dots + c_1 x + c_0)
\end{align*}
for some $c_{N-2}, \dots, c_0$ with $c_0 = \sum_{i=0}^{N-1}x_i$ due to the induction hypothesis.  Comparing coefficients of the $N$-th and $0$-th power of $x$ this polynomial has the form
\begin{align*}
x^{N} + d_{N-1}x^{N-1} + \dots + d_1 x + d_0
\end{align*}
with $d_0 = - x_N \cdot c_0 = x_N \cdot x_{N-1} \cdots x_0$, since $-x_N=x_N$ over $\GF(2^m)$.
\item[(2)]  For arbitrary $1\leq j \leq N$ it is
\[
	p(x_j) = \prod_{i=1}^N (x_i-x_j) + \prod_{i=1}^N x_i = \prod_{i=1}^N x_i.
\]
\item[(3)] Set $\prod_{i=1}^N x_i = y$ and assume $x \neq x_1, \dots, x_N$ with $p(x)=y$.  The polynomial $p(x)-y$ now has $N+1$ zeroes but degree $\leq N$, which means it must be that $p(x)=y$, which contradicts that $p$ has degree $N$.
\end{description}
\end{proof}
\end{lemma}

The following recognizer, illustrated in Figure~\ref{fig:serial}, works as follows: inputs are first shortened with a hash function.  The result is then fed into a polynomial constructed as in the previous Lemma.
The construction uses the following parameters:
\begin{itemize}
	\item[N] Number of elements to be recognized.
	\item[n] Length of elements to be recognized ($\mathcal{U}=\bits^n$).
	\item[q] Maximum number of phishing attempts.
	\item[m] Length of fingerprint ($\FP=\bits^m$).
\end{itemize}

\begin{definition}
\label{def:recogdb}
Let $q,n,m,N$ be natural numbers with $q,m<n$ and $1<N<2^m$.  
Set $\univ = \bits^n$.  
Define the recognizer $\Pi=(\univ, N, \init, \test)$ with $\init$ and $\test$ as follows.

\begin{description}
	\item[Init]
		on input $M=\{x_1, \dots, x_N\}$ with $|M|=N$, choose $\db \randsel \GF(2^n)^{q+N}$ 
		then calculate $\hat{x}_i := h_{\db}(x_i)$ and set
	\[
		p(\xi) := \prod_{i=1}^{N} (\xi-\hat{x}_i) + \prod_{i=1}^{N} \hat{x}_i.
	\]
	Let $a_{N-1}, \dots, a_1$ be the coefficients of this polynomial, set $k=(a_{N-1}, \dots, a_1)$, 
	and return $\langle \db, k, \test(\db, k, x_1) \rangle$.
	\item[Test]
		on input $\db, k$ and $x$, 
		parse $k$ as $k=(a_{N-1}, \dots, a_1)$ and set $p(\xi) := \sum_{i=1}^{N-1} a_i \xi^i$, 
		calculate $\hat{x} = h_{\db}(x)$.  Return $p(\hat{x})$.
\end{description}
One can easily verify that indeed $\test(\db, k, x) = \prod_{i=1}^{N} \hat{x}_i$ for $x \in \{x_1, \dots, x_N\}$, so the correctness requirement is satisfied.
\end{definition}

Since the database $\db$ is chosen independently of the set $M$ this recognizer is trivially secure against disclosure.  Now consider security against collision.  The idea of the proof is that since $H$ is strongly $(N+q)$-universal, the random variables $h_{\db}(x)$ for $x \in Q$ are uniformly distributed and independent.  Since $|h_{\db}(M)| \leq N$ the chance that $h_{\db}(x) \in h_{\db}(M)$ for one particular $x\in Q$ is thus $\leq N \cdot 2^{-m}$, and the chance that this holds for any one $x \in Q$ is given by repeating this experiment $q$ times.

\begin{theorem}
\label{theo:wdb}
Definition~\ref{def:recogdb} is a $(q, \infty, \varepsilon)$-secure recognizer with
\begin{align*}
	\varepsilon &= 1- (1 - N/2^m)^q.
\end{align*}

\begin{proof}
Let $M=\{x_1, \dots, x_N\}$ the set of recognized elements and $Q=\{x_{N+1}, \dots, x_{N+q}\}$ the set of queries.  Since $h$ is $(N+q)$-universal the random variables $h_{\db}(x_1), \dots, h_{\db}(x_{N+q})$ are all uniformly distributed over $\GF(2^m)$ and independent. The chance that for one fixed $i$ it is $h_{\db}(x_{N+i}) \in h_{\db}(M)$ is $N/2^m$.  The chance that for any $i$ it is $h_{\db}(x_{N+i}) \in h_{\db}(M)$ is thus $1- (1-N/2^m)^q$.
\end{proof}
\end{theorem}


\section{Prototype Implementation}
\label{sec:impl}

We now report details of our prototype implementation.  We implemented the recognizer from 
Definition~\ref{def:recogdb}
together with a visual hash calculation as a browser extension for the Tor Browser.  We now briefly explain how it works.

In order to initialize a new recognizer, users open the extensions' settings page and are requested to enter the onion domains  they want to recognize.  
The extension then generates a database, password and fingerprint combination.  The database is stored in the persistent storage of the extension, and the password and a fingerprint visualization are both shown to the user.  The password is held in volatile memory only, and users have to enter it again once they restart their browser.

To verify a domain name the extension presents itself as a toolbar button, shown in Figure~\ref{fig:extensioncheck}.  
The popup features a check button, that when clicked, calculates the fingerprint of the currently visited domain name, and visualizes it as visual hash.  We consider the study and design of appropriate visual hashing schemes future work.  For our prototype, we picked the Mosaic Visual Hash~\cite{fietkau_2019_Using}
as a placeholder because it featured a freely available javascript implementation.\footnote{\url{https://github.com/jfietkau/Mosaic-Visual-Hash} Accessed: 2023-09-12.}

Calculation of a fingerprint involves evaluating a polynomial of degree $\approx 100$ over $\GF(2^{256})$, which takes about 4.5 seconds on a MacBookPro15,2 with 2.4 GHz Intel Core i5 using Tor Browser Version 12.5.4 and an unoptimized implementation.  Storage usage is negligible.  We share more details in appendix~\ref{sec:vardetails}.

In the remainder of this section we explain selected details of the implementation: how passwords are generated and input errors handled, and the recognizer's parameters.

\subsection{Passphrase and Entry Errors}
We encode recognizer keys ($k$) in a series of words chosen from a wordlist.  
We explain the details of our wordlist in~appendix~\S\ref{sec:wordlist}.
It can encode 10.5 bit per word, and two words always have an edit distance of at least three.  This is useful in order to detect input errors, as we will now explain.

\label{sec:pwinput}

A system built from recognizers can not tell when users mistype their password.  If the system had a way of telling a correct from a wrong password, then so can an adversary with access to the same data.  This opens the system up for exhaustive attacks on the password.
But when users enter a wrong password, the fingerprint calculation is incorrect, leading to falsely classified items.

A similar problem exists for loss-resistant password managers~\cite{wolf_2012_IMobileSitter,bojinov_2010_Kamouflage,guldenring_2015_Knock}.  Previous work~\cite{wolf_2012_IMobileSitter,bojinov_2010_Kamouflage}
suggested a visual indicator telling users about a wrong password, or stored abridged checksums for probabilistic checks~\cite{guldenring_2015_Knock}.
We followed a different direction.  Since the words chosen by our encoding all have (at least) edit distance three, this allows us to detect (at least) two input errors per word.  When users enter their password, our implementation continuously tests if the entered words, separated by hyphens, appear in the wordlist and signal an error if not.

\begin{table}
\centering
\caption{Summarizes suitable parameters for our recognizer for $N$ elements to achieve a security of $\varepsilon$ in the collision game with $q$ phishing attempts for a key size of $|k|$ bit. The two rightmost columns show the number of bits of the fingerprint that users must recognize, and the number of words they must to recall.}
\label{table:allconstructions}
\begin{tabular}{ccc|cc|cc}
 $N$ & $q$ & $m$ & $\varepsilon$ & $|k|$ & \# Bit Fingerprint & \# Words \\
\hline
 2 & 100 	& $21$ 	& $9.5\cdot 10^{-5}$ & 21 & 21  & 2 \\
 3 & 100 	& $21$ 	& $1.4\cdot 10^{-4}$ & 42 & 21  & 4 \\
 4 & 100 	& $21$ 	& $1.9\cdot 10^{-4}$ & 63 & 21  & 6 \\
 5 & 100 	& $21$ 	& $2.3\cdot 10^{-4}$ & 84 & 21  & 8
\end{tabular}
\end{table}

\subsection{Security Parameters}
\label{sec:parameters}
Implementing the recognizer from Definition~\ref{def:recogdb} requires to select suitable parameters.  
Among them are the parameters $q$ and $\varepsilon$ bounding adversaries' success probabilities.

We assume users are willing to accept a chance of $\varepsilon := 3\cdot 10^{-4}$ that our protection mechanism might fail.  This is the same chance of guessing a 4-digit PIN in three tries, which appears to be an acceptable risk for someone guessing one's credit card PIN.  
We do not know how often users might encounter phishing sites.  In lack of a better estimate we guess that defending against $q := 100$ phishing attempts might be a reasonable choice.  
We do not have to pick a time limit $t$ for adversaries, since our construction from \S\ref{sec:polyrec} is secure for unbounded adversaries.
In summary, we want our recognizer to be $(100, \infty, 3 \cdot 10^{-4})$-secure. 

Comparing these numbers with the formula in Theorems~\ref{theo:wdb} we can now determine the missing parameter $m$ for several choices of $N$.
Table~\ref{table:allconstructions} shows that $m=21$ achieves our goal for $N$ up to $5$.

\subsection{Discussion and Limitations}
\label{sec:recoggaps}
Provided users are able to properly recognize recognizer fingerprints, our recognizer protects users in line with our security parameters against phishing.  A proper evaluation requires to study if users are indeed able to do so, though, and we leave this as future work.

Recognizers still have some limitations that we now discuss.  First, our discussion of recognizers does not consider 
changes to the recognized set of elements.  Designing recognizers and user interactions that allow for changes is
an interesting challenge:  recognizers necessarily lose information about the stored set $M$, and given only $k$ and $\fp$, it is not possible to fully determine $M$.
Second, our construction can not detect when one recognized domain impersonates another recognized domain.   
One way to address this is by feeding a hash $H(t||d)$ of page title $t$ and domain name $d$ into the recognizer.  This sacrifices unconditional security against collisions, though.
We leave the exploration and study of solutions to these problems for future work.

Finally, the key length and thus passwords grows linearly with the number of recognized items $N$.  
It quickly approaches an upper limit where an encryption-based approach appears more appealing than a recognizer (e.g.\ 80 bit).  Five items are enough to reach beyond that upper limit.


\section{Related Work}
\label{sec:relatedwork}
We now review related work that we did not already discuss in~\S\ref{sec:mainAcademicCountermeasures} and Appendix~\S\ref{sec:academicCountermeasures}.

\paragraph{Hash visualization}
Hash visualization approaches, discussed in detail in \S\ref{sec:keyvis} and \S\ref{sec:keyvis2}, compute a visual fingerprint for a key.  Users are expected to recognize the fingerprint correctly in order to detect a phishing attack.

Recognizers have two advantages compared with plain hash visualization in past approaches:  recognizers map multiple inputs to the same fingerprint and produce a much shorter fingerprint.  We expect that shorter fingerprints simplify the creation of human distinguishable fingerprints, since less information must be mapped onto distinguishable images. 

\paragraph{Phishing in the Tor Network}
Biyukov~\cite{biryukov_2013_Trawling} previously enumerated onion service addresses.  They briefly mention onion service addresses that they suspect to be either backup or phishing sites targeting the silk road marketplace.

Winter~\cite{winter_2018_How} studied how users understand onion services.  They discovered that phishing users as well as operators consider phishing to be a threat.  Utilizing DNS root data, they discovered that users attempt to visit onion domains that are slightly different than the original domain, suggesting that phishing attacks may happen.  
In their study they also evaluated how users visit onion domains.
Users reported to copy and paste onion domains, click links they encounter, using bookmarks or Google, or type domain names from their notes.  They conclude that:
{\begin{quote}Given the high number of (possibly insecure) home-baked solutions, a Tor Browser extension that solves the problem of saving and tracking onion links seems warranted.\end{quote}}

Yoon~\cite{yoon_2019_Doppelgangers} crawled onion sites in order to detect phishing sites.  They crawled 28,928 HTTP Tor hidden services, and reported 901 unique phishing domains.
Barr-Smith and Wright~\cite{barr-smith_2020_Phishing} also crawled onion sites in order to detect networks of phishing sites.
More recently, Steinebach~\cite{steinebach_2021_Phishing} crawled onion pages and evaluated simple heuristics to detect phishing pages.  

Independent and parallel to our work, Wang et.\ al.\ investigated the security mechanisms of onion markets~\cite{wang_2024_Analysis}.  Their work discusses various security mechanisms, such as warrant canaries, and also discusses some phishing protection mechanisms.  Compared to their work, our work focusses on phishing protection mechanisms and provides a more comprehensive and in-depth discussion.

\paragraph{Probabilistic Data Structures}
Our work also relates to approximate membership query data structures (AMQs).  AMQs like Bloom filters~\cite{fan_2014_Cuckoo} 
store set-membership information approximately.  

AMQs in adversarial scenarios have been the subject of recent works.   Naor and Eylon formalized the security of Bloom filters in an adversarial model~\cite{naor_2015_Bloom,naor_2019_Bloom}.  They consider adversaries that attempt to degrade their performance by crafting clever inputs.  Their work considers constructions to be adversarial resilient when they retain their error probability in the face of an adversary with oracle access to the bloom filters' state.
Clayton et.\ al.~\cite{clayton_2019_Probabilistic} expanded on this work.  Among other things, they consider security against adversaries that get access to the AMQs state.  They also briefly mention that an ``interesting direction is to consider what information data structures leak via their public representations''.
Filić et.\ al.~\cite{filic_2022_Adversarial} investigate this in their privacy formalizations.  
Our work relates to theirs in that our privacy notion considers information theoretic security, while 
theirs considers time-bounded adversaries.

\section{Conclusion and Future Work}
We investigated phishing protection mechanisms in use by onion services and discussed phishing countermeasures previously
proposed in the literature.  
Building on the idea of hash visualization we defined recognizers:  a password-protected fingerprint calculation mechanism where multiple domain names give the same fingerprint, and that hides the set of recognized items.  We provide a prototype
implementation as browser extension for the Tor web browser, filling in a gap previously formulated by Winter et.\ al.~\cite{winter_2018_How}.

We see several areas for future work.  Our work motivates the study of visualization for short fingerprints suitable for recognition.  
The shorter length compared to previous proposals gives hope to finding robust visualizations.
The design and interaction with mutable recognizers is another particularly interesting challenge:  Recognizers 
necessarily lose information about recognized items, which complicates user interaction.

Another target for study are the tradeoffs between recall and recognition:  In a recognizer for $N$ elements users remember one password and recognize one fingerprint.  It might turn out that users actually prefer to not remember any passwords at all, but prefer to
recognize $N$ fingerprints instead.  Our recognizers motivate to study users' preferences and 
performance in these tasks.

\bibliographystyle{ACM-Reference-Format}
\bibliography{main}

\appendix

\section{Implementation Details}

\subsection{Various Implementation Details}
\label{sec:vardetails}
The recognizer needs to store a database (see Definition~\ref{def:recogdb}).  Using the parameters in Table~\ref{table:allconstructions}, the database consists of 104 numbers of length $n$-bit, where $n$ is the size of the items to be recognized.  The public key of an onion service is a Curve25519 public key of $256$ bit~\cite{goulet_2013_Tor}, which means $n=256$.  In total this sums up to less than 5 kilobytes.

For the implementation of the finite field $\GF(2^{256})$ we use the irreducible polynomial $1+ x^{121} + x^{178} + x^{241} + x^{256}$ as calculated by Živković~\cite{zivkovic_1994_Table}, and for the finite field $\GF(2^{20})$ we use the irreducible polynomial $1 + x^3 + x^5 + x^{20}$ as calculated by Zierler and Brillhart~\cite{zierler_1968_Primitive}.
For the visual hash we use the Mosaic Visual Hash~\cite{fietkau_2019_Using} and it's open source implementation on GitHub.\footnote{\url{https://github.com/jfietkau/Mosaic-Visual-Hash/}, Accessed: 2023-09-12}

\subsection{Wordlist}
\label{sec:wordlist}
We used the EFF short wordlist 2.0 from~\cite{bonneau_2016_EFF} as basis.  We removed the word \texttt{yo-yo} for containing a hyphen, and added 154 words taken from the EFF large wordlist in~\cite{bonneau_2016_EFF}.  This results in 1449 words, allowing to embed 10.5 bit of information per word.  
Note that contrary to the original short wordlist 2.0, entries in our wordlist do not have a unique three-letter prefix.

We now explain how we selected the added words.  Towards this let $d(a,b)$ denote the Levenshtein distance~\cite{levenshtein_1966_Binary} between two strings $a$ and $b$, and for a set $S$ let $d(a,S) = \min \{d(a,x) : x \in S \}$, and $d(S)= \min \{d(a, b) : a,b \in S \wedge a \neq b\}$.  Let now $S$ be the short wordlist 2.0 without \texttt{yo-yo} and $L$ the large wordlist.

We first removed all entries $x$ from L where $d(x, S)<3$.  We then removed all entries $x$ with $d(x, L \setminus \{x\}) < 3$.  This gives us the set $L'$ with
\[
	L' = \{x \in L : d(x,S)\geq 3 \wedge d(x, L \setminus \{x\}) \geq 3\}.
\]
Note that now 
\begin{enumerate}
	\item $d(S) \geq 3$,  due to how $S$ is constructed~\cite{bonneau_2016_EFF},
	\item $d(L') \geq 3$, and
	\item $\forall x \in S, y \in L': d(x, y) \geq 3$.
\end{enumerate}
Any $x,y \in S \cup L'$ with $x \neq y$ is covered by one of these cases, so we get $d(S \cup L') \geq 3$.

We then further removed all entries from $L'$ with length greater than 4.  This resulted in 214 words, and we picked the following 154 by random choice:
icy,
afoot,
album,
alibi,
aloha,
amaze,
amiss,
among,
anew,
annex,
anvil,
april,
aqua,
askew,
await,
banjo,
bleep,
bless,
blitz,
bluff,
borax,
boxer,
briar,
brim,
buddy,
cadet,
cameo,
chemo,
cocoa,
comfy,
crux,
dares,
debug,
decoy,
denim,
derby,
ditch,
dowry,
ducky,
ebay,
edgy,
elite,
elope,
email,
evade,
evil,
exact,
expel,
fancy,
ferry,
fetch,
film,
five,
gamma,
gauze,
gawk,
genre,
gents,
gnat,
gonad,
green,
henna,
icy,
idiom,
image,
jawed,
jaws,
jimmy,
jinx,
judo,
jump,
kebab,
kiln,
kitty,
knelt,
knoll,
kudos,
lapel,
legal,
lego,
level,
lurch,
macaw,
magma,
maker,
march,
mauve,
maybe,
mocha,
moody,
mouth,
mumps,
nacho,
nerd,
ninth,
nutty,
okay,
ovary,
panda,
panic,
pesky,
peso,
petri,
photo,
plaza,
pluck,
polio,
posh,
pupil,
radar,
rehab,
remix,
rigid,
rumor,
saga,
saucy,
scuba,
sedan,
sepia,
shaft,
siren,
skype,
slurp,
snub,
stir,
stony,
syrup,
tarot,
thigh,
thud,
tibia,
turf,
tutor,
tutu,
tweak,
twerp,
twig,
ultra,
until,
unwed,
veto,
walk,
wham,
whiff,
whole,
widen,
width,
worry,
wrath,
xbox,
zero,
zesty,
zips,
zit,
zoom.

\section{Supplementary Data}
\label{sec:appendixdata}

\begin{table}
\caption{Shows a full list of categories of onion services in our dataset.}
\label{table:categories2}
\begin{tabular}{cc}
\toprule
Count & Category \\
\midrule
  12 & Cryptocurrency\\
  11 & Technology\\
  11 & Marketplace\\
   8 & Forum\\
   7 & Email\\
   7 & Link Directory\\
   6 & News\\
   6 & Social\\
   4 & VPN\\
   4 & Blog\\
   3 & Search Engine\\
   3 & Postage\\
   3 & Image Board\\
   3 & Hosting\\
   2 & Publication Platform\\
   2 & Government\\
   1 & Wiki\\
   1 & Webpage Archival\\
   1 & Various Services\\
   1 & PGP key directory\\
   1 & Git\\
   1 & Forum / Marketplace\\
   1 & File Sharing\\
   1 & Education\\
   1 & Data Leaks Directory\\
   \bottomrule
\end{tabular}
\end{table}

Table~\ref{table:categories2} shows the full list of onion service categories in our dataset and the number of services
in each category.


\section{Known Countermeasures}
\label{sec:academicCountermeasures}

This section reviews previously proposed phishing countermeasures against our requirements from \S\ref{sec:requirements}: protection against phishing, disclosure and censorship, and independent security against disclosure.  Sometimes
prior academic work suffers from adoption problems, which we consider additionally where necessary.
Note that we evaluate these approaches only with respect to \emph{our}
scenario of applying them to onion services, and that our threat model is likely different than the one
they were originally intended for. 

To identify relevant research works we relied on Franz' recently published survey~\cite{franz_2021_SoK}, that we enriched with our own additional literature research.  
We won't fully present each approach but rather concentrate on the properties relevant for our threat model.  To keep the presentation succinct we also only discuss when a solution fails to provide one of our properties.

\subsection{Identity Indicators}
Users fall for phishing attacks because they mistake the displayed website for the mimicked one and don't pay attention to browser indicators~\cite{dhamija_2006_Why}.  One line of work investigated methods to make users pay more attention to the web browsers' security features or URL bar, e.g.\ ~\cite{lin_2011_Does,miyamoto_2014_EyeBit}.
These approaches don't work for onion services, because it is inherently difficult for users to tell the difference between an original and a phishing onion domain
 (Phishing: \none). 

\subsection{Sitekey and Personalization}
\label{sec:sitekey}
Services might thwart phishing by authenticating themselves towards users, similar to what
we observed in \S\ref{sec:onionLogin}.  Upon entering the user id the website shows the user a 
previously registered secret, e.g. an image as in Sitekey~\cite{bankofamerica_2005_Bank}, or requires
the user to click on the secret among a set of images~\cite{herzberg_2013_Forcing}.
UI dressing~\cite{iacono_2014_UIDressing} by Iacono et.\ al.\ customizes the entire web site using a 
per-user custom background image and layout.
In all these approaches the website sends a static secret to the client, and they are thus 
defeated by sophisticated phishing attacks 
(Phishing: \none). 

\subsection{Client Secrets Stored in Cookies}
One alternative is to store information on clients that makes the login process different
on different domains.  In Phorcefield~\cite{hart_2011_PhorceField}, a cookie customizes the login
process for one domain, and phishing pages on another domain will not have access to this cookie.
This requires that browsers store a map between images and websites, though, which leaves traces of the website on users devices (Disclosure: \none;).

Juels et.\ al.\ proposed cache cookies~\cite{juels_2006_Cache}, and Braun et.\ al.\  client-side cookies in PhishSafe~\cite{braun_2014_PhishSafe} to defend against phishing.  
As in PhorceField, browsers' security policies enforce that phishing pages don't have access to cookies. Browsers, again, need to store a map between website and secret, leaving users vulnerable to disclosure attacks
(Disclosure: \none;).

Though not using cookies, device fingerprinting (see~\cite{alaca_2016_Device} for a discussion) roughly falls in the same category. 
But since the Tor Browser strives to remove any kind of fingerprintability~\cite[\S 2.2]{TorDesign_2018} this is not applicable to onion services.

The approaches above utilize that it is difficult for one domain to steal the cookies of another.  Two-Factor authentication, which we discuss next, similarly utilizes additional client secrets that are difficult to steal.

\subsection{Two-Factor Authentication}
\label{sec:2fa}
Another approach is to give clients additional secrets that are difficult to steal with phishing.
We discern between two different 2FA methods: ones that verify the identity of the
website requesting authentication, we call those dynamic, and ones that do not, we call those static.
Wo don't discuss 2FA relying on secondary channels, such as SMS or email, since this might compromise users' anonymity and are thus not applicable to onion services.

\subsubsection{Static Two-Factor Authentication} 
The time-based one-time password~\cite{rfc6238} (TOTP) generates a short lived login token from a shared secret and the current time, which users submit on login.  The tokens are just as easily phished as other authentication credentials, and modern phishing kits\footnote{\url{https://github.com/kgretzky/evilginx2}, Accessed: 2023-09-12.}  support to intercept them (Phishing:~\none). TOTP relies on a shared secret between website and user, which makes a users' security against disclosure depend on the website. (Independent:~\none).  
Though TOTP implementations regularly store the websites where TOTP secrets are used, this is technically not necessary (Disclosure: \full).

\subsubsection{Dynamic Two-Factor authentication}
\label{sec:dyn2fa}

BeamAuth~\cite{adida_2007_Beamauth} embeds a secret token in a bookmark that must be clicked during the login process.  Login attempts at phishing pages redirect to the correct page, protecting against phishing.  The bookmark encodes the original domain, though (Disclosure: \none).

Gajek et.\ al.~\cite{gajek_2008_Provably} adapted the TLS protocol to support mutual authentication between web pages and users, called Browser-based Mutual Authentication (BBMA). 
After authenticating clients' computers with TLS client certificates, BBMA authenticates web pages by
showing a secret image as in Sitekey.  But since servers must be able to authenticate client certificates, BBMA can never provide independence (Independence: \none).
In Let's Authenticate~\cite{conners_2022_Let}, a central Certificate Authority issues on-demand certificates to users wishing to login to a website.  Clients don't need to store an account database because the CA keeps an encrypted list of domain names and account identifiers.  The website and CA both observe account identifiers upon each login, though.  Furthermore, encrypted data stored with the CA is protected by a physical authenticator, such as a FIDO2 token.   (Independence: \half; Disclosure: \half)

Modern 2FA approaches, such as those built on top of WebAuthn~\cite{Webauthn}, use challenge response protocols to verify the identity of the website prior to authentication.   
In WebAuthn public key credentials contain a reference to the domain where credentials are used~\cite[\S 4 Public Key Credential Source]{Webauthn}.  Credentials are suggested to be either stored on the authenticator device~\cite[\S 6.2.2]{Webauthn}, breaking disclosure.  Or they are suggested to be stored at the website~\cite[\S 6.2.2]{Webauthn}, encrypted using a private key embedded
in the device, which would break independence (Disclosure: \half; Independent: \half).

Essentially challenge response protocols must ensure that requests cannot be relayed by man-in-the-middle adversaries.
So clients must be able to verify that a challenge truly originates from the visited domain. 
As long as the client can verify this, a disclosure-adversary can abuse this to identify visited domains.
A different class of cryptographic protocols avoids this problem, which we discuss next.

\subsection{Mutual Authentication Protocols (PAKE)}
\label{sec:pake}
Password authenticated key exchange (PAKE) protocols mutually authenticate users and websites using a previously shared secret: the users' password.  
Modern protocols protect against sophisticated phishing attacks, and the early SRP protocol~\cite{wu_1998_Secure} was even added to TLS~\cite{taylor_2007_Using}.  
PAKE protocols fully solve the phishing problem as long as users and websites share a secret.
But PAKE protocols have never been adopted by web browsers~\cite{engler_2009_It}.

One of the reasons  PAKE hasn't found adoption is that it requires a trusted user interface for password input.  UI spoofing attacks~\cite{tygar_1996_WWW,depaoli_1997_Vulnerability,tygar_1996_WWW} that simulate a login window bypass PAKE and can capture credentials~\cite{engler_2009_It}.
Short: PAKE protocols requires a trusted path~\cite{dod_1985_orange}.

\subsubsection{Trusted Path}
\label{sec:trustedpath}
Several suggestions established a trusted path between the user and the browser or web page.  Unfortunately, most of them are rather invasive in the user experience.

Ye and Smith proposed to visualize trusted browser windows using animated window borders that blink synchronized with a trusted reference window~\cite{ye_2005_Trusted}.  
Dhamija and Tygar suggested Dynamic Security Skins (DSS)~\cite{dhamija_2005_Battle}, customizing
the background of trusted input elements with background images, and utilizing a trusted browser
area to display reference images.

Jakobsson and Myers proposed to augment password-entry with a character-by-character 
feedback mechanism, called Delayed Password Disclosure (DPD)~\cite{jakobsson_2007_Delayed}.
Upon each key press, users have to verify if the correct feedback image is shown.

Another line of work uses secure attention sequences.  WebWallet~\cite{wu_2006_Web}, which we discuss
further below (\S\ref{sec:credtracking}), requires users to press a key sequence in oder to submit sensitive data.  
Spoofkiller~\cite{jakobsson_2012_Spoofkiller} conditions users to press the power button
in order to login.  

All proposals share that they are quite invasive in the user experience, and none of them have found
adoption in browsers.
 
\subsection{Trusted Path Captchas}
\label{sec:obc}
In a separate line of work Captchas establish a trusted path between a website and the user.  The idea is similar to how Captchas are used in some onion services in our dataset (see~\S\ref{sec:apcaptchas}): 
The server encodes parts of its public key in a Captcha.  When users solve the Captcha within a trusted browser component, the browser uses the solution to verify the connection.

Saklikar et.\ al.~\cite{saklikar_2008_Public} proposed to encode excerpts of a websites' public key in a Captcha.  A second
Captcha determines which excerpts are encoded.  
Petrillo et.\ al.~\cite{ferraropetrillo_2014_Design} and Ahmad et.\ al.~\cite{ahmad_2018_Detecting} later improved these ideas. 
Urban et.\ al.~\cite{urban_2017_Riddle} explored tying a Captcha's contents to online banking transactions to protect users against malware infected machines.

All these ideas require that it is possible to generate Captchas that can not be solved by machines.  Recent advances in image recognition and the continuing difficulty in generating new Captchas that resist attacks raise doubts that this approach is future-proof.  Captchas do not work against human-in-the-middle attacks, of course, and do not help against unsophisticated phishing attacks (Phishing: \half).

\subsection{Security Toolbars}
Web browsers can use heuristics, allow-, or deny-lists\footnote{Older papers refer to these terms as white- and blacklisting. Today allow- and deny-listing are considered to be more appropriate terms.} in order
to detect phishing pages and warn their users.  These are the most widely used approaches in practice today, and for example, all major web browsers support the Google Safe Browsing denylist~\cite{google_2023_Google}.

\subsubsection{Allow- and deny-lists}
Allow- and deny-lists (e.g. \cite{google_2023_Google}) have to be curated, which opens them up for censorship attempts~\cite{biddle_2023_Apple}
(Censorship: \none). 
They are also necessarily always outdated, offering limited protection against phishing (phishing: \half).  

\subsubsection{Heuristics}
Heuristics are able to detect unknown phishing sites.
Chou et. al.~\cite{chou_2004_Clientside} proposed SpoofGuard, that uses multiple stateless and stateful heuristics to calculate a spoof index for a website.  Stateless heuristics examine the URL and images on the web site.  But since attackers know these heuristics, they can design their sites to bypass them (Phishing: \half).  

Stateful heuristics make use of the browsing history, previously entered user names and passwords, and submitted data.  These heuristics leave users vulnerable to disclosure (Disclosure: \none), while they too offer only  limited protection against phishing (Phishing: \half).  
AuntieTuna~\cite{ardi_2016_AuntieTuna} uses personalized lists of previously visited web sites, and
falls in the same category.

\subsubsection{Assisted Decision Making}
Another approach is to help users decide if a visited page is a phishing site.  In iTrustPage, users are assisted with automated web searches~\cite{ronda_2008_Itrustpage}, and  BayeShield~\cite{likarish_2009_BayeShield} uses a conversational user interface.  
Users already visit phishing pages mimicking onion services after searching for the correct domain.  We thus don't think these approaches substantially add to what users already do
(Phishing: \none).

\subsubsection{Credentials Tracking}
\label{sec:credtracking}
AntiPhish~\cite{kirda_2006_Protecting} by Kirda and LoginInspector by Yue~\cite{yue_2012_Preventing} monitor where users enter login information.  They detect when users enter their credentials for one site at another site, and raise an alarm.  Both require to store a mapping of credentials to website, encrypted in AntiPhish and hashed in LoginInspector, making both susceptible to disclosure attacks.

WebWallet~\cite{wu_2006_Web} prevents users from entering sensitive data on web pages, and provides a separate user interface instead.
Web Wallet utilizes a trusted path using a secure key sequence.  When submitting sensitive data, Web Wallet checks interactively 
with the user if the destination website is allowed to receive it.

Both AntiPhish and Web Wallet protect against phishing, but require to storage sensitive data, exposing users to a disclosure attack 
(Disclosure: \none).

\subsection{Password Managers}
Password managers can prevent phishing attacks by ensuring that credentials are never sent to wrong sites.  There are essentially two ways this can be done: credentials undergo some destination-specific transformation (password generator), or the password manager verifies the identity of a site for users.  

\subsubsection{Password Generators}
Password generators, first proposed as part of the Janus system by Gabber et.\ al.~\cite{gabber_1997_How} and later improved~\cite{ross_2005_Stronger,halderman_2005_Convenient},  
calculate site-passwords from a master password and some auxiliary information.  The site-password calculated for a phishing page will be different from the
one calculated for the correct page.  

A brute force attack on the master passwords can, however, calculate
the master password from the site-password (compare \S\ref{sec:requirements}) .
Password generators don't require trusted third parties and they are not susceptible against disclosure attacks.  But generated passwords are related to one another, making them susceptible to phishing (Phishing: \half).

\subsubsection{Password Managers}
Browser-based password managers ensure that passwords are only auto-filled on the correct website.
The password manager must know a mapping of passwords to domains, making it vulnerable to disclosure.  
(Disclosure:~\none).

\subsection{Mitigating Attacks Against Encryption}
Theft-resistant passwords managers utilize ambiguity to protect site-passwords when master passwords are not strong enough for regular encryption~\cite{bojinov_2010_Kamouflage,wolf_2012_IMobileSitter, chatterjee_2015_CrackingResistant,guldenring_2015_Knock}.
They don't intend to hide the map between passwords and websites, though, making them susceptible to disclosure attacks
(Disclosure: \none).

One alternative is to encrypt the entire database in a fashion that slows down adversaries.  One line of work incorporated human work in the encryption process.  Canetti et.\ al.~\cite{canetti_2006_Mitigating,daher_2008_POSH} suggested to use Captchas to slow down adversaries attacking disk encryption,  and Blocki suggested to incorporate inkblot images~\cite{blocki_2013_GOTCHA}.  Both assume that humans can perform image recognition tasks that machines can't.

Pepper~\cite{manber_1996_Simple} and Key stretching~\cite{rfc2898} make brute force attacks on passwords more expensive, but cannot prevent them.  They also require authentication delays of over one second in order to be effective ~\cite{guldenring_2015_Knock,blocki_2018_Economics}
(Disclosure: \half).

\subsection{Domain Names for Onion Services}
One line of work proposed name resolution systems or other ways to discover onion domains.
In Sauteed Onions~\cite{dahlberg_2022_Sauteed}, onion domains are embedded in certificates of non-onion domains to allow monitoring in certificate transparency logs.  Discovering attacks does not protect
against phishing, though 
(Phishing: \none, Censorship:~\none).

Onion names~\cite{onionnames} currently allow onion sites running the SecureDrop software to register a  memorable onion domain name, for example, \url{theguardian.securedrop.tor.onion}.  
Onion names aid discovery and memorability of onion domains.  But they do not directly protect against phishing (Phishing: \half), and the curated nature of onion names leaves them open to censorship 
(Censorship:~\none).

OnionDNS runs an unmodified DNS server over Tor~\cite{scaife_2018_OnionDNS}, but relies on
an anonymous trusted third party operating the service, making it vulnerable to censorship and phishing 
(Phishing: \none; Censorship:~\none).

The Onion Name System~\cite{victors_2017_Onion}, a distributed name system for onion domains, uses a proof of work based lottery system to limit the effects of land rushing and domain squatting attacks, which it can not fully prevent, though.
Name resolution systems ultimately guide discovery and provide memorability of onion domains.  In itself they don't authenticate the service to users, thus offering limited protection against phishing  (Phishing: \half).

Tor proposal 194~\cite{sai_2012_Mnemonic} 
suggest a mnemonic system to encode onion domain names in a sentence.  Mnemonic encoding improves the memorability of onion domain names, but users still need to recall the same amount of information as without mnemonic encoding (280 bit encoded as 56 characters for each domain name) 
(Phishing:~\none).

\subsection{Petnames}
\label{sec:petnames}
Petname systems bind a cryptographic identity to a local name.  As Yee noticed in~\cite{yee_2006_Passpet} they are basically an address book.  Using petnames with certificates is, in a way, Ellisons's suggestion on an alternative to the CA system~\cite{ellison_1996_Establishing}, as later formalized by Rivest and Lamport in the SDSI system~\cite{rivest_1996_SDSIa}.
Petnames have also recently been suggested in the context of onion services,\footnote{\url{https://gitlab.torproject.org/tpo/applications/tor-browser/-/issues/40845}, Accessed 2023-11-29.} but we are unaware of any particular
proposal on how to implement the mapping of petnames to onion domains.

Close proposed the YURL~\cite{close_2004_Trust} petname system.  YURL stores a local map associating public key hashes with petnames.  An improved system~\cite{close_2006_Petname} stored petnames alongside TLS certificate information in a bookmark.  
Yee and Sitaker~\cite{yee_2006_Passpet} suggested a combination of petnames with password generators, named Passpet.  Petnames in Passpet are associated with the domain name or TLS certificate information.
Herzberg and Jbara~\cite{herzberg_2008_Security} proposed Trustbar, extending Close's petnames by allowing users to choose logos as well.  As an alternative to choosing Logos, the authors also consider logos included in TLS certificates, as have previously also been proposed by Jøsang et.\ al.~\cite{josang_2001_Authentication} and specified in RFC 3709~\cite{rfc3709}.

All the aforementioned petname systems require a local mapping of petnames to public keys, which makes them vulnerable to disclosure attacks (Disclosure: \none).

\subsection{Hash Visualization}
\label{sec:keyvis2}
Perrig and Song proposed to present public key information as a visual hash to users~\cite{perrig_1999_Hash}.  In follow up works Tay~\cite{tay_2004_Visual} and Dabas~\cite{dabas_2005_Browser} suggested to use that to visualize web site certificate information.  
OpenSSH 5.1 introduced a visual fingerprint calculation, coined drunken bishop by Loss et.\ al.~\cite{loss_2009_Drunken}.  
These ideas were later also picked up in a discussion\footnote{\url{https://lists.torproject.org/pipermail/tor-dev/2015-August/009302.html}, Accessed 2023-11-29.} among Tor developers, where it was considered to visualize
onion domains.  
All these works share that visual fingerprints encode a large number of bits. 
To the best of our knowledge, how well humans recognize these visual fingerprints was never studied in
the context of phishing attacks 
(Phishing:~?).

In what probably comes closest to our work, Markham suggested in a blog post a similar idea to thwart typo squatting: adding a 12 bit hash value of a domain name to the domain~\cite{markham_2005_Phishing}.

\end{document}